\documentclass[osajnl,twocolumn,showpacs,superscriptaddress,10pt]{revtex4-1} 
\usepackage{amsmath,amssymb,graphicx}
\usepackage[colorlinks,linkcolor=blue,citecolor=blue,urlcolor=blue,anchorcolor=blue]{hyperref}
\usepackage{color}
\begin{document}

\title{Interactions of incoherent {localized beams} in a photorefractive medium}

\author{Yiqi Zhang}
\email{zhangyiqi@mail.xjtu.edu.cn}
\affiliation{Key Laboratory for Physical Electronics and Devices of the Ministry of Education \&
Shaanxi Key Lab of Information Photonic Technique,
Xi'an Jiaotong University, Xi'an 710049, China}

\author{Milivoj R. Beli\'c}
\email{milivoj.belic@qatar.tamu.edu}
\affiliation{Science Program, Texas A\&M University at Qatar, P.O. Box 23874 Doha, Qatar}

\author{Huaibin Zheng}
\author{Haixia Chen}
\author{Changbiao Li}
\author{Jianeng Xu}
\affiliation{Key Laboratory for Physical Electronics and Devices of the Ministry of Education \&
Shaanxi Key Lab of Information Photonic Technique,
Xi'an Jiaotong University, Xi'an 710049, China}

\author{Yanpeng Zhang}
\email{ypzhang@mail.xjtu.edu.cn}
\affiliation{Key Laboratory for Physical Electronics and Devices of the Ministry of Education \&
Shaanxi Key Lab of Information Photonic Technique,
Xi'an Jiaotong University, Xi'an 710049, China}

\begin{abstract}
  We investigate numerically interactions between two bright or dark incoherent {localized beams} in an {strontium barium niobate} photorefractive crystal
  in one dimension, using the coherent density method.
  For the case of bright beams, if the interacting beams are in-phase, they attract each other during propagation and form
  bound breathers; if out-of-phase, the beams repel each other and fly away.
  The bright incoherent beams do not radiate much and form long-lived well-defined breathers or quasi-stable solitons.
  If the phase difference is $\pi/2$, the interacting beams may both attract or repel each other,
  depending on the interval between the two beams, the beam widths, and the degree of coherence.
  For the case of dark incoherent beams, in addition to the above the interactions also depend on the symmetry of the incident beams.
  As already known, an even-symmetric incident beam tends to split into a doublet, whereas an odd-symmetric incident beam tends to split into a triplet.
  When launched in pairs, the dark beams display dynamics consistent with such a picture and in general obey soliton-like conservation laws,
  so that the collisions are mostly elastic, leading to little energy and momentum exchange. But they also radiate and breathe while propagating.
  In all the cases, the smaller the interval between the two interacting beams, the stronger the mutual interaction.
  On the other hand, the larger the degree of incoherence, the weaker the interaction.
\end{abstract}


\maketitle 

\section{Introduction}
Incoherent spatial solitons in photorefractive (PR) media have come into research focus and attracted much attention in the last two decades \cite{mitchell_prl_1996,mitchell_prl_1997,mitchell_nature_1997,christodoulides_prl_1997,christodoulides_ol_1997,nayyar_josab_1997,
christodoulides_prl_1998a,christodoulides_prl_1998b,coskun_ol_1998,chen_science_1998,akhmediev_prl_1998,shkunov_prl_1998,
carvalho_pre_1999,coskun_pre_1999,sukhorukov_prl_1999,krolikowski_pre_1999,
kip_science_2000,coskun_prl_2000,krolikowski_pre_2000,coskun_ol_2000,
christodoulides_pre_2001,huang_oc_2007,zhang_cpl_2009,zhang_cpb_2009}. Lately, many types of incoherent solitons were
reported both in theory and experiment. This remarkable progress not only opened up new research fields
in soliton science and nonlinear optics, but also broadened the potential applications of optical solitons.

The progress was facilitated by the appearance of new methods for the treatment of incoherent localized beams in PR media:
the coherent density method \cite{christodoulides_prl_1997,christodoulides_ol_1997,coskun_ol_1998,coskun_pre_1999},
the self-consistent multimode method \cite{mitchell_prl_1997,christodoulides_prl_1998a,christodoulides_prl_1998b,carvalho_pre_1999,
coskun_prl_2000,akhmediev_prl_1998,krolikowski_pre_2000,sukhorukov_prl_1999},
and the mutual coherent function method \cite{nayyar_josab_1997,shkunov_prl_1998}.
They were developed independently to describe exactly such sorts of solitary waves in theory,
however it was soon demonstrated that these three seemingly different theoretical methods are
equivalent to each other in inertial nonlinear media \cite{christodoulides_pre_2001}.
Recently, the coherent density method seems to be employed more often than the other in papers
on incoherent solitons \cite{coskun_ol_2000,huang_oc_2007,zhang_cpb_2009,zhang_cpl_2009}.
Thus, incoherent bright solitons interacting with incoherent dark solitons or coherent dark solitons (which would enhance
the spatial coherence) have been investigated with the coherent density method \cite{coskun_ol_2000}.
In addition, interactions of incoherent solitons were also considered in Refs.  \cite{huang_oc_2007, ku_prl_2005}.
However, there are still topics related with the interactions of copropagating incoherent solitons that are worth investigating.
For example, interactions of dark incoherent beams have never been discussed before. This is done here.

{Thus far, interactions of bright coherent solitons have been investigated thoroughly \cite{stegeman_science_1999}.
Fusion, fission, and repulsion of soltions have been reported.
On the other hand, interactions of dark coherent solitons \cite{thurston_josab_1991,torres-cisneros_jmo_1995,ohta_arxiv_2010}
have been less investigated, yet are quite different from those of bright solitons.
From this point of view, interactions of incoherent dark solitons may show interesting phenomena and deserve further attention.}

In this paper, we first introduce the coupled nonlinear Schr\"odinger-like integro-differential equations
that provide a general description of the incoherent solitary waves in PR media. Then, we numerically
solve them by employing the coherent density method, combined with the beam propagation method.
Following this procedure, we investigate the interactions between incoherent bright and dark solitary waves, respectively.
The target medium used in our numerical simulations is the biased strontium barium niobate (SBN) PR crystal,
which is often employed in soliton experiments \cite{chen_ol_1997,krolikowski_prl_1998,krolikowski_apb_1999,krolikowski_prl_2000,terhalle_pra_2012}.
{We refrain from calling these localized beams solitons, because they sometimes display
inelastic collisions, breathe and radiate, but propagate quasi-stably over considerable lengths that outdistance
typical crystal thicknesses.}

The paper is  organized as follows: in Sec. \ref{model}, we briefly introduce the theoretical model;
in Sec. \ref{results}, we discuss the interactions of bright and dark incoherent beams in detail;
in Sec. \ref{conclusion}, we conclude the paper.

\section{Theoretical Model}
\label{model}

We assume that the beams propagate along the $z$-axis and are allowed to diffract along the $x$-axis.
Under these conditions and according to the theory developed in Refs. \cite{christodoulides_ol_1997,coskun_ol_2000},
in a biased SBN PR crystal with optical $c$-axis oriented in the $x$-direction,
the two incoherent light beams evolve according to the equation
\begin{align}\label{eq1}
i\left( {\frac{{\partial f}}{{\partial z}} + \theta \frac{{\partial f}}
{{\partial x}}} \right) & + \frac{1}{{2k}}\frac{{\partial ^2 f}}{{\partial x^2 }} \notag\\
& - \frac{{k_0 }}{2}n_e^3 r_{33} E_0 (1+\rho) \frac{f}{{1 + I_f (x,z)}} = 0,
\end{align}
from an initial condition at $z=0$
\begin{equation}\label{eq2}
f(z = 0,x,\theta ) = \sqrt{r_f G_N(\theta )}\phi _0 (x),
\end{equation}where
\begin{equation}\label{eq3}
I_f (x,z) = \int_{ - \infty }^{+\infty}  {\left| {f(x,z,\theta )} \right|^2 d\theta }
\end{equation}
is the total beam intensity.
In Eqs. (\ref{eq1})-(\ref{eq3}), $ f $ represents the so-called coherent density of the coupled incoherent beams
and $\theta $ is the ratio $k_x/k$ of the transverse to the longitudinal wave number.
In the paraxial propagation we consider, this ratio is approximately equal to the angle
of the beam with respect to the $z$-axis. In addition,
$k=2\pi n_e /\lambda _0 $ (in which $\lambda_0$ is the free-space wavelength and
$n_{e}$ is the extraordinary refractive index of the SBN crystal),
$ r_{33}$ is the electro-optic coefficient involved in the generation of the space-charge field,
$\rho$ the intensity when $x\to\pm\infty$, and
$E_0 =\pm V/W$ is the external bias field needed for the development of nonlinearity
(here $V$ is the applied bias and $W$ is the transverse width of the SBN crystal).
For a biased SBN:75 crystal, the typical values of the parameters mentioned above are
$n_{e}$=2.3, $r_{33}$=1022 pm/V, $W$=6 mm, and $\lambda_{0}$=488 nm \cite{christodoulides_prl_1997,coskun_ol_1998},
which we are going to use in our simulations.

Concerning the initial condition -- that is, the incident beam -- the function
$G_{N}$({$\theta $}) is the normalized angular power spectrum of the incoherent source,
$\phi_{0}(x)$ is the input complex spatial modulation function of the two incoherent beams, and
$r_{f}$ is the maximum intensity of the incoherent beams.
We assume that the normalized angular power spectrum of the incoherent source is Gaussian,
i.e., $G_N(\theta )=\exp (-\theta ^2/\theta _0^2 )/(\sqrt \pi \theta _0 )$
\cite{christodoulides_prl_1997,christodoulides_ol_1997,coskun_ol_1998,coskun_pre_1999},
where $\theta_0$ is the angular half-width of the power spectra associated with the incoherent beams.
It also a measure of the incoherence -- the larger $\theta_0$, the more incoherent the beams.
As concerns the modulation function $\phi_0$, for bright incoherent beams we define it as
\begin{equation}\label{eq4}
\phi _0 (x) = \exp \left( - \frac{(x-d)^2}{2x_0^2} \right) + \exp(il\pi) \exp \left( - \frac{(x+d)^2}{2x_0^2} \right),
\end{equation}
in which $2d$ is the distance between the two Gaussian input beams, $l$ determines the phase difference between the beams,
and $x_0$ is related to the full width at half maximum (FWHM) of the Gaussian beam.
Corresponding to Eq. (\ref{eq4}) we have \[x_0 = \frac{\rm FWHM}{2\sqrt {\ln 2} }.\]

For dark incoherent beams, we define $\phi_0$ as
\begin{equation}\label{eq5}
\phi _0 (x) = 1 + \tanh\left(\frac{x-d}{x_0}\right) - \tanh\left(\frac{x+d}{x_0}\right),
\end{equation}
for the odd-symmetry case.
In Eq. (\ref{eq5}), \[x_0=\frac{\rm FWHM}{\ln \left( 3+2\sqrt{2} \right)}~ {\rm and}~ d \geq d_{\rm th}=\frac{\ln3}{2}x_0,\]
to guarantee the two solitons are really dark. For the even-symmetry case, we use
\begin{subequations}\label{eq6}
  \begin{equation}
  \phi _0 (x) = \sqrt{1- \frac{S(x)}{\max\{S(x)\}}},
  \end{equation}
with
  \begin{equation}
  S(x) = {\rm sech}^2\left(\frac{x-d}{x_0}\right) + {\rm sech}^2\left(\frac{x+d}{x_0}\right),
  \end{equation}
\end{subequations}
in which \[x_0=\frac{\rm FWHM}{\ln \left( 3+2\sqrt{2} \right)}.\]

{We should mention at this point that the analytical incoherent soliton solution is hard to obtain, owing to the complexity of the nonlinear dynamical problem. The
variational methods \cite{malomed_book_2002,buccoliero_prl_2007,shen_oc_2009,kong_ol_2010,shen_pra_2010,pu_oc_2012} can be used to solve for the asymptotic soliton solutions;
however, as usual with variational methods, they are as good as the presumed ansatz solutions are -- optimized but still approximate analytical solutions.
Considering that we focus on the interacting dynamics of incoherent bright and dark beams in detail, which cannot be treated analytically,
the incident beams given in Eqs. (\ref{eq4})-(\ref{eq6}) are still similar to the incidents used in previous literatures \cite{christodoulides_prl_1997,christodoulides_ol_1997,coskun_ol_1998},
and are good enough for our purpose.
The interactions of asymptotic incoherent solitons obtained by using the variational method are beyond the scope of this paper.}

\section{Numerical simulations and discussions}
\label{results}

\subsection{Bright incoherent beams}

\begin{figure}[htbp]
\centering
  \includegraphics[width=\columnwidth]{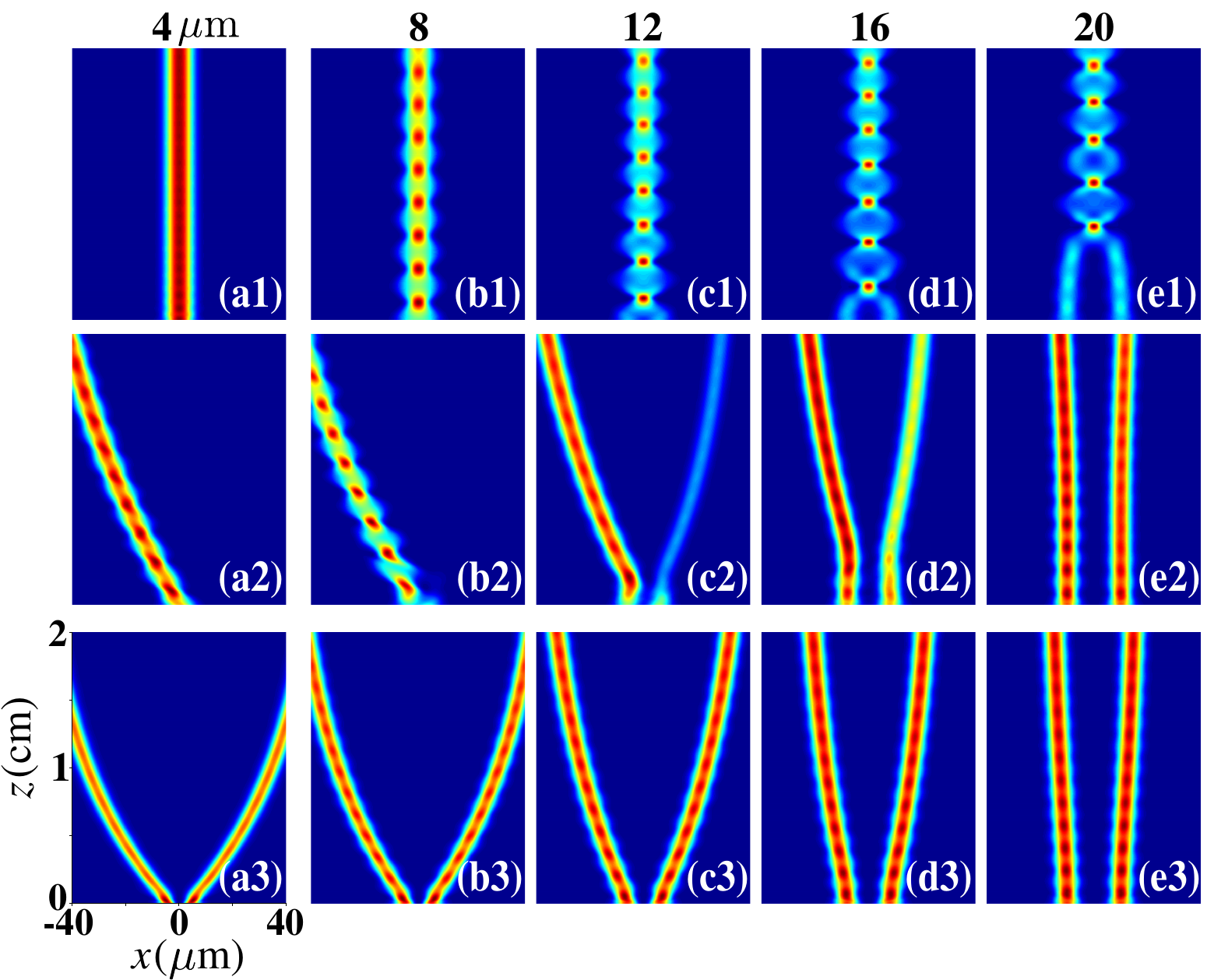}
  \caption{Interaction of two incoherent bright beams with different intervals $2d$ between them (the numbers displayed on the top of the panels).
  Other parameters are $l=0,~0.5$, and 1 for (a1)-(e1), (a2)-(e2), and (a3)-(e3), respectively, ${\rm FWMH}=7~ \mu {\rm m}$,
  $V =550$ V, and $\theta_0=1$ mrad.}
  \label{fig1}
\end{figure}

We first investigate the interactions of two in-phase (viz. $l=0$ in Eq. (\ref{eq4})) incoherent beams;
the numerical results are shown in Figs. \ref{fig1}(a1)-\ref{fig1}(e1).
It is seen that two incoherent solitons attract each other during propagation and form a bound breather.
With the interval between beams increasing, the interaction strength decreases
and the two incoherent beams need a longer distance to form a breather.

If $l=1/2$, the phase difference between the interacting incoherent beams is $\pi/2$
and the corresponding results are displayed in Figs. \ref{fig1}(a2)-\ref{fig1}(e2).
It is interesting to note that the interacting beams still attract each other,
to form a deflected breather if the interval is not big, as shown in Figs. \ref{fig1}(a2) and \ref{fig1}(b2).
When the interval is big enough, the interacting beams repel each other, as shown in Figs. \ref{fig1}(c2)-\ref{fig1}(e2).
The reason is that there exist attraction and repulsion simultaneously between the two interacting beams;
when the interval is small, the attraction prevails,
while if the interval is big, the repulsion prevails.

The reason why the breather is deflected is also not difficult to explain.
From Eq. (\ref{eq4}) we can find the transverse group velocity $v_x$ of the wave packet, which can be written as
\begin{align}\label{eq7}
v_x = &-\frac{x-d}{x_0^2}\exp \left( - \frac{(x-d)^2}{2x_0^2} \right) \notag \\
&- \exp(il\pi) \frac{x+d}{x_0^2} \exp \left( - \frac{(x+d)^2}{2x_0^2} \right).
\end{align}
When $d$ is not big, the two beams behave as one wave packet, so
that we can check the speed at $x=0$, to obtain $v_x|_{x=0}\neq 0$.
Hence, the breather should deflect.
Furthermore, the beams accelerate transversely,
as there are forces acting on them.

In the out-of-phase case {(viz. $l=1$ in Eq. (\ref{eq4}))}, exhibited in Fig. \ref{fig1}(a3)-\ref{fig1}(e3),
the interaction is predominantly repulsion.
And the smaller the interval, the stronger the repulsion.
Also note that in the interactions of bright incoherent beams, practically no
radiation is observed -- the solitons form fast from the incident beams
and continue to propagate quite stably.

\begin{figure}[htbp]
\centering
  \includegraphics[width=\columnwidth]{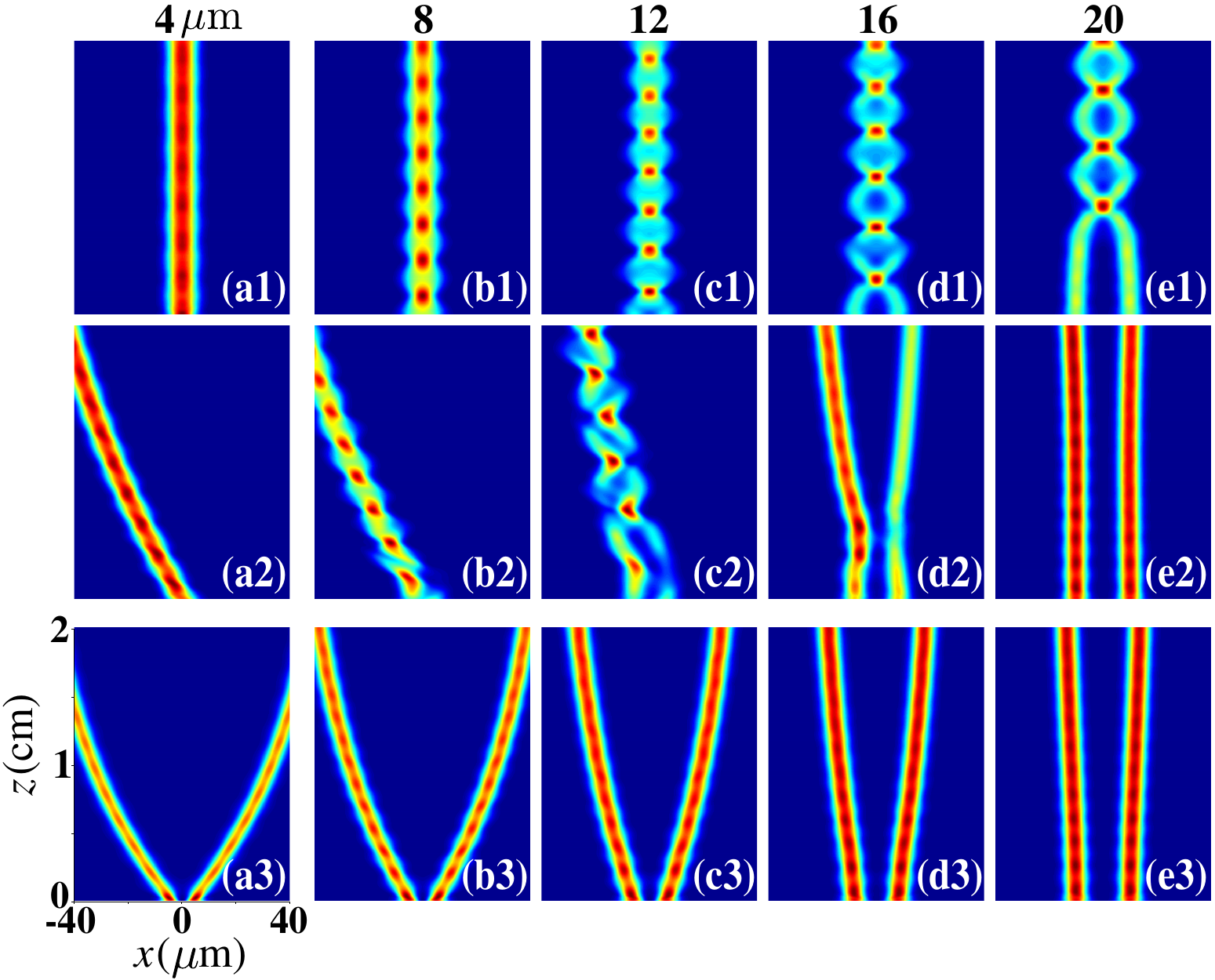}
  \caption{Same as Fig. \ref{fig1}, but with $\theta_0=3$ mrad.}
  \label{fig2}
\end{figure}

To check the influence of the degree of incoherence, which is estimated by the width of the
power spectrum, we change $\theta_0$ from 1 mrad to 3 mrad and redo the simulation;
the corresponding results are displayed in Fig. \ref{fig2}.
In comparison with those shown in Fig. \ref{fig1}, we find that the interactions weaken.
Especially, by comparing Fig. \ref{fig2}(c2) with Fig. \ref{fig1}(c2),
we see that the competition between attraction and repulsion is also affected by $\theta_0$, the degree of incoherence.
It can be predicted that the bigger the $\theta_0$ and the smaller the interval makes the attraction the more dominant.

\subsection{Dark incoherent beams}

As established previously \cite{coskun_ol_1998,zhang_cpb_2009,zhang_cpl_2009},
dark incoherent beams tend split into a doublet under the even conditions,
or into a triplet or evolve into a gray soliton under the odd conditions
(depending whether FWHMs are large or small, respectively).
We discuss the interactions of dark incoherent beams under the odd conditions first, and then under the even conditions.
To these ends, we use the  modulation functions displayed in Eqs. (\ref{eq5}) and (\ref{eq6}), respectively.

\begin{figure}[htbp]
\centering
  \includegraphics[width=\columnwidth]{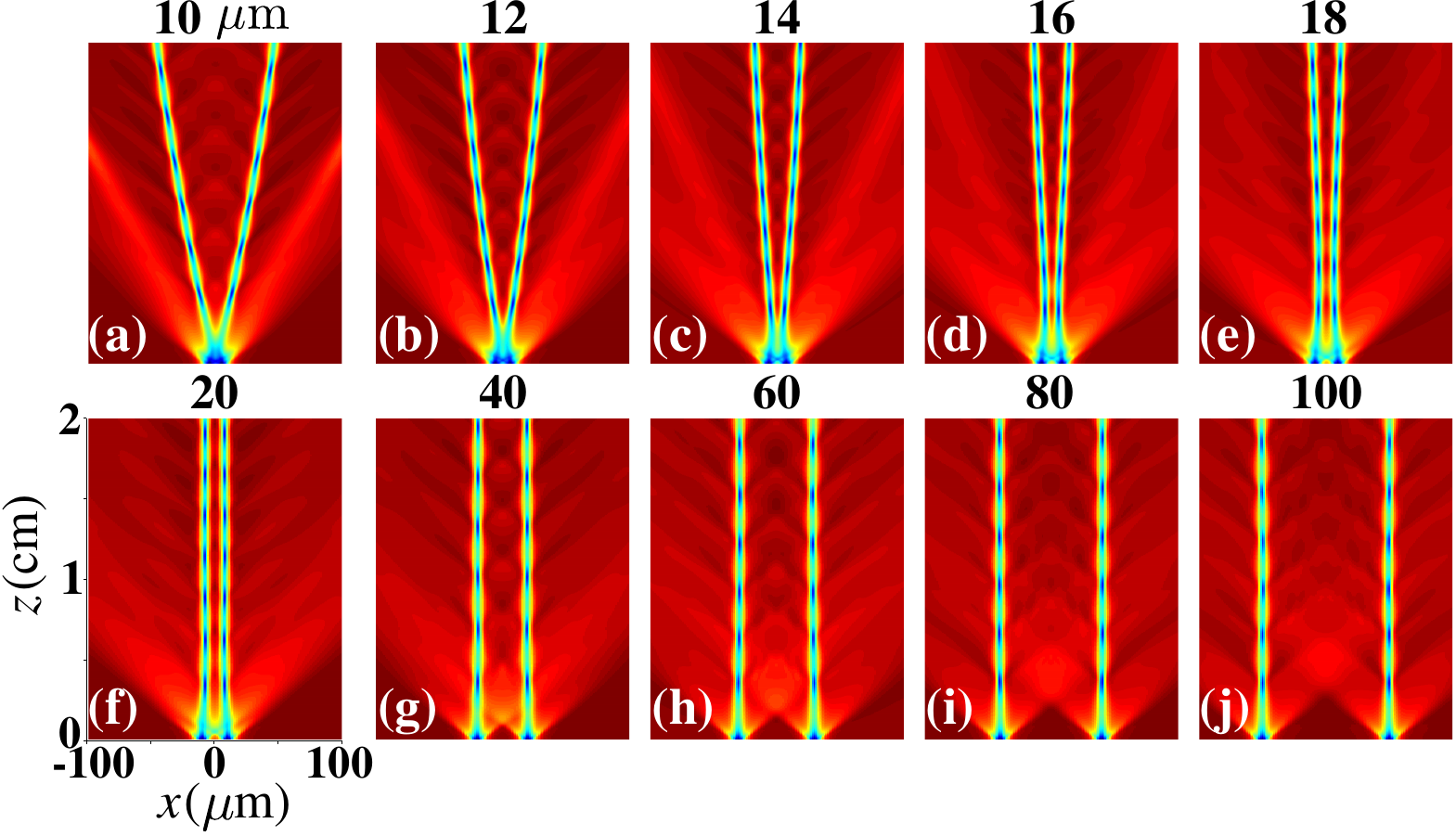}
  \caption{Interaction of two dark incoherent beams with ${\rm FWMH}=10~ \mu {\rm m}$,
  for changing interval $2d$ under odd conditions.
  Other parameters: $V =-550$ V and $\theta_0=9.6$ mrad.}
  \label{fig3}
\end{figure}

We assume first that the incident beams are narrow, ${\rm FWMH}=10~ \mu {\rm m}$; the results are shown in Fig. \ref{fig3}.
In Figs. \ref{fig3}(f)-\ref{fig3}(j), in which the intervals are bigger than the FWHM,
one can see that two gray solitons are formed, between which there is little or no interaction.
When $d-d_{\rm th}$ is small, as shown in Figs. \ref{fig3}(a)-\ref{fig3}(e),
there is attraction first and then the repulsion between the two interacting dark incoherent beams.
The reason is that the two dips in the input incoherent beams are so close that they will fuse into one at the beginning,
strongly radiate, and then separate into two gray solitons.
In addition, the function displayed in Eq. (\ref{eq5}) looks like an even function for small intervals,
so the interaction is then in analogy with the even case,
which leads to the formation of a doublet \cite{coskun_ol_1998}.
One should note that, generally, the dark beams breathe and initially strongly radiate but exchange little energy as they propagate.

\begin{figure}[htbp]
\centering
  \includegraphics[width=\columnwidth]{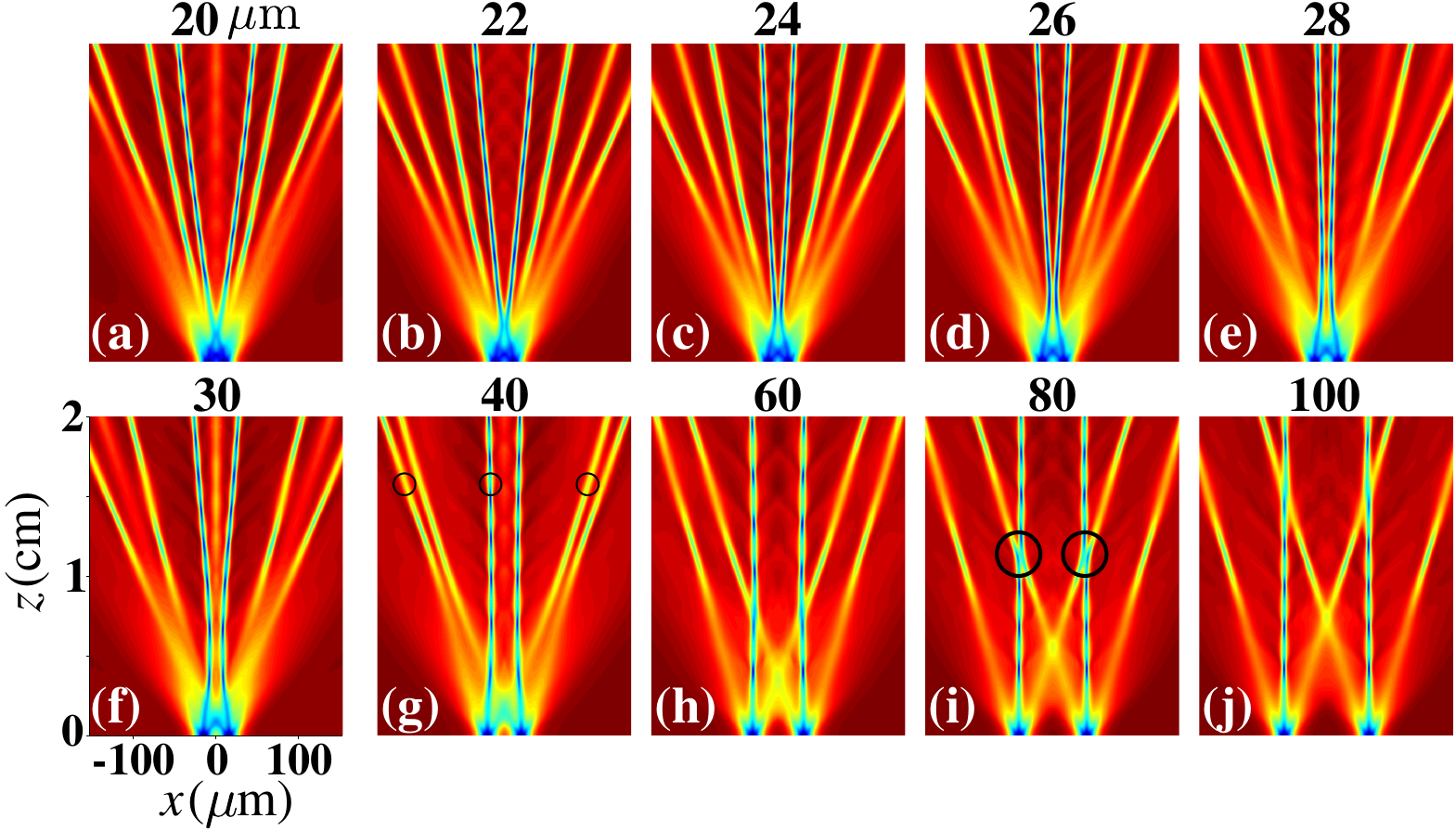}
  \caption{Same as Fig. \ref{fig3} but with ${\rm FWMH}=20~ \mu {\rm m}$.}
  \label{fig4}
\end{figure}

When the FWHM is increased to $20~ \mu {\rm m}$, the results are displayed in Fig. \ref{fig4}.
For $d$ close to $d_{\rm th}$, as shown in Figs. \ref{fig4}(a)-\ref{fig4}(d), the dynamics is complex.
There are many channels generated during propagation,
and the two main channels in the middle first attract and then repel each other.
The reason is similar to that of the case in Fig. \ref{fig3} --
the input is similar to an even case with a much wider FWHM,
which leads to additional channels \cite{zhang_cpb_2009,zhang_cpl_2009}.

However, when the interval is large, the input cannot be in analogy to the even case and has to revert to the odd case.
This means that only six channels can form during propagation -- three beam components for each of the dark incoherent beams --
and the ensuing interaction is between the two triplets.
This is clearly seen in Figs. \ref{fig4}(e)-\ref{fig4}(j).
In Fig. \ref{fig4}(g), the three channels marked by the three black circles originate from the same dark incoherent beam source:
the main channel close to the center and the two secondary channels.
One can note that the secondary channels are not symmetrically distributed about the main channel.
For this case, the interaction is relatively strong
and we can also find that the secondary channels from different source attract each other.
Note the strong radiation formed in the central region, as well as the strong radiation emanating sideways in the beginning.

In Fig. \ref{fig4}(i), with the interval increasing further,
the interaction between the two main channels becomes weak.
In addition, we see that the secondary channels collide with the main channels.
As marked by the two circles, the collision does not lead to the energy exchange between the secondary and main channels,
but leads to some momentum exchange.
The reason for this elastic collision is that the intensities of the two participating beams are almost of the same size at the collision point.
However, in the case of Fig. \ref{fig4}(j), there are also collisions between the secondary and main channels,
but these collisions seem to be transparent, because the intensity of the secondary channel is much smaller than that of the main channel.
We can also observe such transparent collisions in Figs. \ref{fig4}(g) and \ref{fig4}(h).
These phenomena obey the usual soliton conservation laws, but with the caveat of considerable initial energy loss in the form of shed radiation.

\begin{figure}[htbp]
\centering
  \includegraphics[width=\columnwidth]{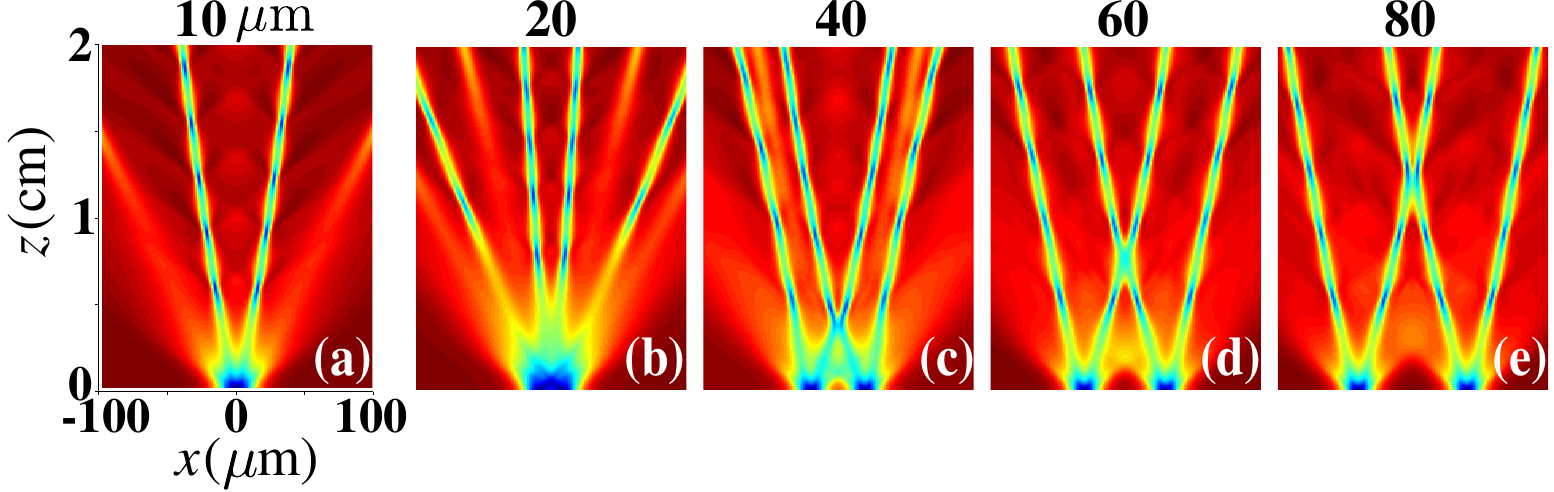}
  \caption{Figure setup is as Fig. \ref{fig4} but under even conditions.}
  \label{fig5}
\end{figure}

Finally, we turn to the cases of even conditions, which are exhibited in Fig. \ref{fig5}.
In Figs. \ref{fig5}(a) and \ref{fig5}(b), the intervals between the input beams are not big, so the total incident beams
can be viewed as hyperbolic-like wave packets with ${\rm FWMH}=30~ \mu {\rm m}$ and ${\rm FWMH}=40~ \mu {\rm m}$, respectively.
Thus, we can classify the case as a splitting of one dark incoherent beam with large FWHM into a doublet,
rather than the interaction of two dark incoherent beams \cite{zhang_cpb_2009}.
Indeed, we observe that more channels begin to appear at longer propagation distances.
In Figs. \ref{fig5}(c)-\ref{fig5}(e),
the doublets from different incidences begin to interact.
Due to the symmetry of the intensity distribution,
the intensities of the two interacting participants are the same,
and the collisions are always elastic. All the formed dark solitary beams breathe and radiate.

\section{Conclusion}
\label{conclusion}

In summary, we have investigated the interactions of bright and dark incoherent localized beams in a PR medium.
For the bright incoherent beams, the interaction is attraction if the interacting beams are in-phase,
and repulsion if they are out-of-phase.
If the phase difference is $\pi/2$, the interaction will either lead to a deflected bound breather or two repulsive solitons,
depending on $d$ and $\theta_0$. The bright beams may breathe or propagate steadily over large distances, without visible radiation.

For the dark incoherent beams, we have discussed both the cases of odd and even symmetry conditions.
Under the odd conditions, the collisions may be elastic or transparent,
while under even conditions, the collisions are always elastic. Quite complex beam interaction
scenarios may exist, but consistent with the theory developed.
These interactions comply with the soliton-like conservation laws, even though the beams breathe and radiate.

\section*{Acknowledgement}
{This work was supported by the 973 Program (2012CB921804), CPSF (2014T70923, 2012M521773),
NSFC (61308015, 11104214, 61108017, 11104216, 61205112),
and NPRP 6-021-1-005 project from the Qatar National Research Fund (a member of the Qatar Foundation).}
%


\bibliographystyle{osajnl}
\bibliography{refs}

\end{document}